\documentclass[lettersize,journal]{IEEEtran}
\usepackage{graphicx}          
\usepackage{amsmath}
\usepackage{tabularx}
\usepackage{amsfonts}
\usepackage{url}
\usepackage[table,xcdraw]{xcolor}
\usepackage{cite}
\usepackage{verbatim}
\usepackage{times,epsfig}
\usepackage{makecell}
\usepackage{psfrag}
\usepackage{subfigure}
\usepackage{stfloats}
\usepackage{setspace}
\usepackage{amssymb}
\usepackage{multirow}
\usepackage{colortbl}
\usepackage[justification=centering]{caption}
\usepackage{fancyhdr}
\usepackage{utfsym}
\usepackage{makecell}
\usepackage{booktabs}
\usepackage{hyperref}
\usepackage{tabularx}

\usepackage{enumitem}
\newlist{steps}{enumerate}{1}
\setlist[steps, 1]{label = Step \arabic*:}
\newcommand*\emptycirc[1][1ex]{\tikz\draw (0,0) circle (0.1);} 
\newcommand*\fullcirc[1][1ex]{\tikz\fill (0,0) circle (0.1);}

\begin{document}
	
\pagenumbering{arabic}

\title{Toward Mixture-of-Experts Enabled Trustworthy Semantic Communication for 6G Networks}

\author{Jiayi He*, Xiaofeng Luo*, Jiawen Kang, Hongyang Du, Zehui Xiong, Ci Chen,\\ Dusit Niyato, \textit{Fellow, IEEE}, Xuemin Shen, \textit{Fellow, IEEE}

\thanks{J. He, X. Luo, J. Kang, and C. Chen are with the School of Automation, Guangdong University of Technology, Guangzhou 510006, China. H. Du and D. Niyato are with the College of Computing and Data Science, Nanyang Technological University, Singapore 639798, Singapore. Z. Xiong is with the Pillar of Information Systems Technology and Design, Singapore University of Technology and Design, Singapore 487372, Singapore. X. Shen is with the Department of Electrical and Computer Engineering,
University of Waterloo, Waterloo, ON N2L 3G1, Canada.

* means equal contribution. (\textit{Corresponding author: Jiawen Kang.})}
}
\maketitle
\pagestyle{headings}

\begin{abstract}
Semantic Communication (SemCom) plays a pivotal role in 6G networks, offering a viable solution for future efficient communication. Deep Learning (DL)-based semantic codecs further enhance this efficiency. However, the vulnerability of DL models to security threats, such as adversarial attacks, poses significant challenges for practical applications of SemCom systems. These vulnerabilities enable attackers to tamper with messages and eavesdrop on private information, especially in wireless communication scenarios. Although existing defenses attempt to address specific threats, they often fail to simultaneously handle multiple heterogeneous attacks. To overcome this limitation, we introduce a novel Mixture-of-Experts (MoE)-based SemCom system. This system comprises a gating network and multiple experts, each specializing in different security challenges. The gating network adaptively selects suitable experts to counter heterogeneous attacks based on user-defined security requirements. Multiple experts collaborate to accomplish semantic communication tasks while meeting the security requirements of users. A case study in vehicular networks demonstrates the efficacy of the MoE-based SemCom system. Simulation results show that the proposed MoE-based SemCom system effectively mitigates concurrent heterogeneous attacks, with minimal impact on downstream task accuracy.

\end{abstract}


\section{Introduction}

    In the era of 6G, the integration of networks spanning space, air, ground, and sea is anticipated to facilitate instant, efficient, and intelligent hyper-connectivity communications. Semantic Communication (SemCom) emerges as a crucial component in this context, serving as an efficient and intelligent paradigm for future communication~\cite{ZHANGPINGSemantic}. The adoption of an encoder-decoder structure within the SemCom system allows for the transmission of only essential information in source messages, enabling the receiver to accurately reconstruct the original message. This capability allows SemCom to surpass the limitations set by Shannon's theorem, accomplishing communication tasks with minimal bit transmission. Consequently, SemCom is poised to bolster wireless communications in scenarios involving a high volume of immersive communication tasks, such as metaverses, in environments with stringent delay requirements like autonomous driving, and in situations where bandwidth resources are limited, such as maritime communication and remote area communication, especially in the 6G landscape~\cite{NTN_SemCom,wang2023semantic}.
    
    The evolution of Neural Network (NN) models has significantly improved the performance of SemCom systems, enabling them to handle a variety of complex tasks, such as multimodal~\cite{MultiModal} and task-oriented communications~\cite{TaskOriented}. However, this progress also highlights inherent vulnerabilities in SemCom systems, primarily linked to the weaknesses of NN models. One critical vulnerability is their susceptibility to semantic noise attacks, where an attacker injects semantic noise into the communication source, leading to decoded message distortion or degraded accuracy~\cite{SemanticNoise}. The open nature of wireless channels exacerbates this fragility, as it permits the injection of semantic noise into the physical channel~\cite{Physical, RobustSC}. Moreover, the presence of a warden in the communication environment poses additional privacy risks. A warden can monitor active communications between devices and exploit exposed model parameters of the semantic decoder to extract sensitive user information~\cite{covert, evestrop}. These weaknesses can easily be exploited to tamper with critical information or eavesdrop on private information. For instance, in a SemCom-based vehicular network, a well-behaved vehicle broadcasts safety messages to nearby vehicles to enhance driving safety~\cite{VTmigration_privacy}. However, if an attacker tampers with the message by adding semantic noise, surrounding vehicles will receive falsified information. This may cause unnecessary emergency braking, leading vehicles into dangerous situations. Therefore, investigating trustworthy semantic communication in 6G networks is urgently needed.

    There are several approaches to address the aforementioned vulnerabilities. To tackle semantic noise attacks, Kang et al.~\cite{SemanticNoise} and Nan et al.~\cite{Physical} suggest enhancing the robustness of semantic codecs by using adversarial training technology. 
    They incorporate adversarial semantics noise into the source or channel during the training of the semantic codec, which can bolster the system's resistance against semantic noise attacks originating from both the source and the physical channel. To shield communication status from detection by a purposeful warden, Du et al.~\cite{covert} propose that the semantic compression rate can be adjusted to reduce the probability of communication behavior being detected. Furthermore, to safeguard against illegal decoding of the semantic information, Luo et al.~\cite{evestrop} advocate deploying a semantic encryption mechanism. Despite these advancements, the challenge of developing a SemCom system that can effectively counteract multiple security threats simultaneously remains largely unexplored in the evolution toward trustworthy 6G networks.

    The Mixture-of-Experts (MoE) model presents a compelling approach to enhance trustworthy SemCom, particularly in the face of multiple heterogeneous security challenges. By partitioning the model and dataset into multiple specialized subgroups, the MoE framework accommodates tailored processing of diverse inputs~\cite{MoE}. In SemCom, this adaptability is crucial for addressing multiple heterogeneous security threats. The semantic codecs can be composed of multiple experts to resist various security vulnerabilities. To optimize the selection of these experts, a gating network is incorporated into the SemCom system. This network intelligently chooses the most appropriate expert or combination of experts based on the current security landscape and the sender's privacy requirements~\cite{LLM_MoE}. This dynamic selection process ensures that the semantic encoding and defense strategies are both effective and aligned with the specific security needs of the communication, thereby improving overall trustworthiness. By leveraging specialized experts and a strategic gating mechanism, this approach intensifies the ability of SemCom systems to safeguard sensitive information and maintain effective communication.
    
    In response to heterogeneous security and privacy challenges, this article pioneers an MoE-based trustworthy SemCom system optimized to improve security and reliability of 6G networks. The main contributions of this article are as follows:	
	\begin{itemize}
	    \item To meet the trustworthy security and privacy requirements in 6G networks, we discuss the promising applications and derived challenges of SemCom, which motivate our use of MoE to invoke appropriate defense schemes for addressing multiple heterogeneous threats simultaneously.
		\item The proposed MoE-based SemCom system can adaptively select the semantic codec according to user-defined security requirements using a gating network. Additionally, our designed SemCom system is extensible, allowing seamless integration of emerging security experts to achieve sustainable trustworthiness for 6G networks.
		\item We provide a case study of vehicular message sharing to show the practicality of our proposed scheme. It considers multiple heterogeneous wireless communication threats including source-end attacks, channel-end attacks, detection attacks, and eavesdropping attacks. Numerical results show that the proposed SemCom system can achieve high accuracy while meeting 6G security and privacy requirements in the face of multiple attacks.

	\end{itemize}
	
\section{Semantic Communication in 6G Network}	

\subsection{Review of Semantic Communication}
SemCom has emerged as a key enabling technology to meet the high-speed and large-scale information transmission requirements in 6G networks~\cite{ZHANGPINGSemantic}. It concentrates on extracting the semantic contents of the source message at the transmitter and interpreting their meanings at the receiver. By exchanging semantic information at both ends, SemCom enables the reconstruction of source messages or direct execution of downstream tasks with tolerance for transmission errors~\cite{NTN_SemCom}.

The Deep Learning (DL)-based encoder-decoder architecture plays a primary role in the SemCom system. In this system, a DL-based semantic encoder at the transmitter effectively encapsulates the essential meaning behind messages while a DL-based semantic decoder at the receiver reconstructs the transmitted semantic information into an understandable form. Various neural networks can be employed by these DL-based semantic codecs to process different types of semantic information. For example, the Transformer exhibits high proficiency in Natural Language Processing (NLP) tasks and is innovatively used in textual SemCom systems~\cite{DeepSC}. Besides, SemCom building upon other DL models such as Variational AutoEncoders (VAEs), Generative Adversarial Networks (GANs), and diffusion models can extract features from different data modalities like images and audio, thereby enhancing the efficiency of message transmission over wireless channels~\cite{SemCom_secure_survey}.

\subsection{Semantic Communication Applications in 6G Networks}
\begin{figure*}[t]
\centering{\includegraphics[width=0.95\textwidth]{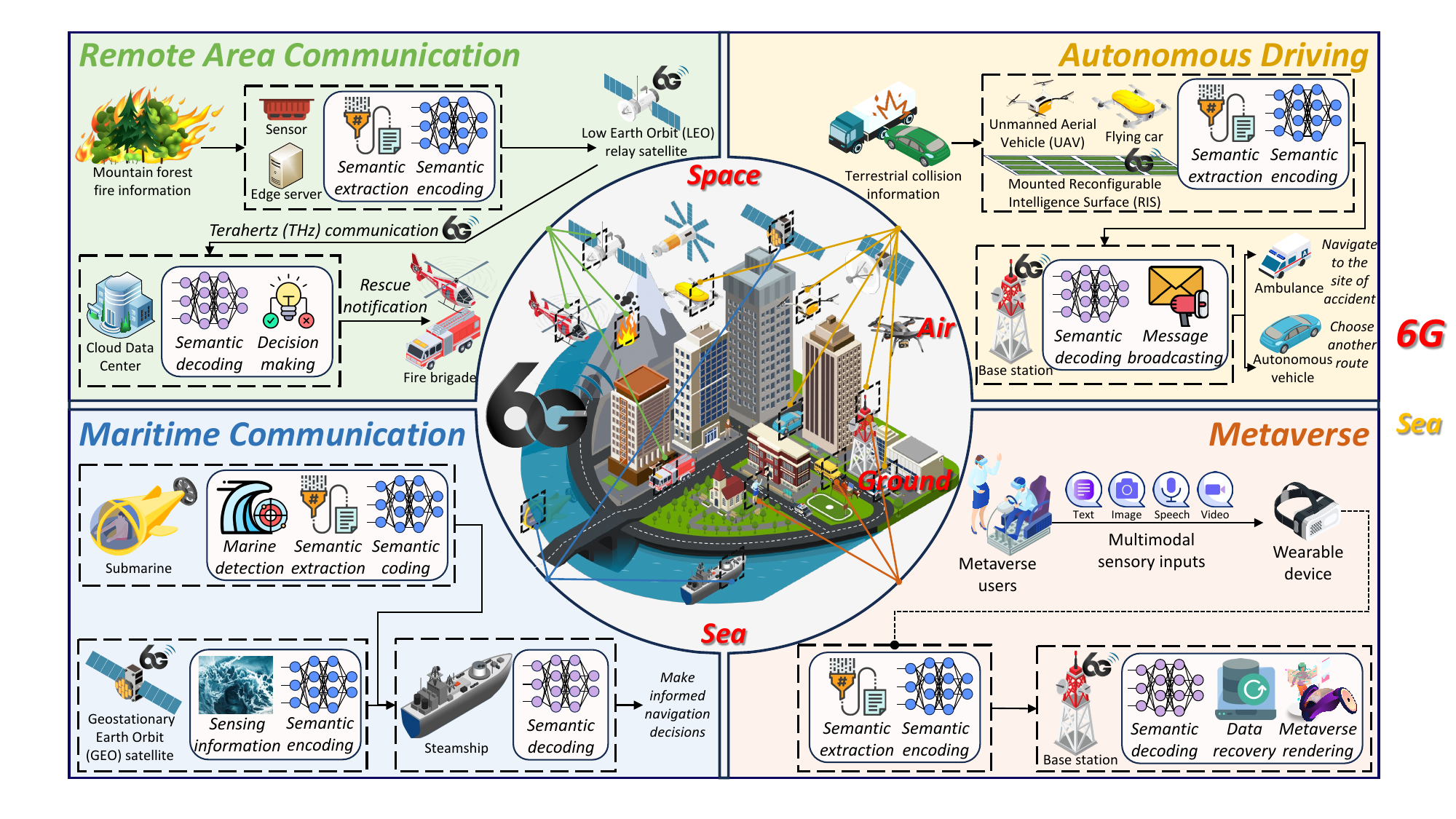}}
\caption{Overview of semantic communication applications in space-air-ground-sea integrated 6G networks.}\label{Fig1}
\end{figure*}

Considering the high efficiency of SemCom, we present four potential application scenarios of SemCom in space-air-ground-sea integrated 6G networks, as depicted in Fig. \ref{Fig1}.

\begin{itemize}
    \item \textbf{Remote area communication:} SemCom can enhance disaster monitoring in remote areas to facilitate emergency response by relevant institutions. For example, if a forest fire occurs deep in the mountain with poor communication conditions, edge servers can extract the essential fire information (e.g., location) detected by the adjacently installed sensor, and then forward it to the neighboring data center after the relay of Low Earth Orbit (LEO) satellite through Terahertz (THz) communication\cite{NTN_SemCom}. The cloud data center can promptly decode the semantic representation to acquire the original complete information and then notify the nearest fire brigade to the scene of the fire.
    \item \textbf{Autonomous driving:} In an intelligent transportation system, autonomous vehicles need to keep connected with other vehicles and roadside infrastructures, e.g., RoadSide Units (RSUs), to enhance contextual awareness of road conditions for safe driving\cite{VTmigration_privacy}. SemCom fosters connected vehicles to share road information with lower latency and higher efficiency. For instance, when low-attitude vehicles, e.g. Unmanned Aerial Vehicles (UAVs), find a traffic accident such as a terrestrial collision during the flight, they can efficiently convey the coded semantic information of the accident, even in urban areas with abundant tall buildings. 

    \item \textbf{Maritime communication:} The harsh oceanic wireless environment challenges maritime communications due to poor channel conditions causing bit distortion\cite{NTN_SemCom}. SemCom mitigates this by conveying the meaning of messages rather than bit streams. For example, a submarine detecting an unknown marine creature with sonar can semantically encode and transmit the information to a targeted vessel (e.g., a steamship) via an acoustic semantic communication system. Meanwhile, a Geostationary Earth Orbit (GEO) satellite can use remote sensing technology to monitor the state of the corresponding sea area and transmit it to the steamship. By decoding both semantic sources, the steamship can comprehensively understand the maritime situation, thereby making informed navigation decisions.
    \item \textbf{Metaverse:} SemCom is expected to play a transformative role in the operation of metaverses~\cite{wang2023semantic, SemanticNoise}. By capturing vital semantic information for transmission, SemCom dramatically reduces communication latency, thereby guaranteeing users' immersive experiences in metaverses. For instance, wearable devices, e.g., VR glasses, can collect multimodal sensory inputs, such as text, image, speech, and video from metaverse users. Due to the limited computing resources of wearable devices, computation-intensive tasks like rendering avatars and virtual spaces should be offloaded to edge servers for further processing~\cite{VTmigration_privacy}. To this end, the raw data are transmitted to the base station via a SemCom system~\cite{SemanticNoise}. With the recovered user data, the base station can render a hyper-realistic metaverse and provide remarkable metaverse services back to users.
\end{itemize}

\subsection{Security Issues and Defense Schemes of SemCom}
As discussed above, SemCom has numerous potential applications that can shape our future lifestyle. Therefore, it is necessary to investigate the security issues and defenses associated with SemCom. As shown in Table~\ref{tab1}, attacks against SemCom can be divided into two categories based on the ultimate goal of attackers, namely, tampering information and eavesdropping information. Tampering can occur at the source end and the channel end, while eavesdropping is divided into two steps: detection of communication status and acquisition of source information. This section introduces these threat models along with existing defense schemes.

\textbf{Tampering Attack from Source End:} By injecting adversarial semantic noise into the source, an attacker can make the semantic decoder fail to restore the source information accurately, leading to incorrect information acquisition at the receiver and erroneous outcomes for downstream tasks. In the white-box mode, the attacker fully grasps the model parameters and weights of the semantic codec, thereby accurately adjusting the gradient information to obtain a minimum semantic noise that causes the semantic codec to produce wrong results. In the black-box mode, the attacker can only query the input and output of the codec a finite number of times to achieve the same goal without gradient information~\cite{SemanticNoise}. Note that a source-end attack does not imply that the sender is the attacker, as an attacker could post noisy images in the real world (e.g., post-it notes on a traffic sign) that the sender unknowingly captures and transmits, thus unexpectedly triggering the attack.

\textbf{Tampering Attack from Channel End:} An attacker can introduce semantic noise into the channel to achieve the same effect as tampering at the source end. However, generating the semantic noise against the channel side requires the attacker to possess more supplementary knowledge. In the white-box mode, the attacker needs to know the model parameters of both the semantic encoder and the semantic decoder, while even in the black-box mode, the attacker is required to be able to query the complete communication input and output. Since tampering attacks add adversarial semantic noise at the source or channel output to achieve their goal, defenders typically assume the presence of a noise generator when training semantic codecs. This adversarial training method ensures that the trained semantic encoders and decoders are minimally affected by semantic noise.

\textbf{Eavesdropping Attack by a Warden:} The goal of a warden is to detect the presence of wireless transmission activity to decipher the location of the communicator or engage in other harmful behaviors. The warden collects signals in the wireless channel at a certain frequency to detect communication activity, which requires the ability to acquire signals in the open wireless channel and analyze the traffic using algorithms. The primary defense scheme is to use covert communication techniques, which involve setting a friendly jammer to emit noise into the communication amplifier to mask the real communication activity. In SemCom, users can also increase the semantic compression rate to reduce the probability of communication being detected~\cite{covert}.

\textbf{Eavesdropping Attack by an Eavesdropper:} An eavesdropper aims to obtain the semantic information transmitted in the channel and correctly decode it to obtain the contents of source messages. This requires the eavesdropper not only to capture the signal in the wireless channel, but also to possess a similar semantic decoder. Using the physical layer technologies, the eavesdropper recovers the source contents with an illegally obtained decoder. Encryption is an effective defense scheme to prevent the eavesdropper from correctly decoding the signal in the channel~\cite{evestrop}. It generates a pair of keys held by the legitimate sender and receiver. The sender mixes the source information with the key and encodes it, and the receiver uses the corresponding key to decode it correctly.

Although there are corresponding defense schemes for each attack, actual communication scenarios often face multiple heterogeneous security threats simultaneously. Existing methods, which include training codecs with specialized techniques, additional encryption operations, or compression operations, cannot simply be merged to address this more complex and realistic scenario. Therefore, it is necessary to explore a new approach that seamlessly integrates various kinds of defense schemes into a SemCom system to support trustworthy 6G networks.

\begin{table*}[t]
\centering
\caption{Comparison of attacks and defenses for semantic communication in 6G networks}
\label{tab1}

\begin{tabular}{|l|p{3.5cm}|p{3.5cm}|p{3.5cm}|p{3.5cm}|}
    \hline
Attack & \multicolumn{2}{p{7cm}|}{\textbf{Tampering}}                                                                                      & \multicolumn{2}{p{7cm}|}{\textbf{Eavesdropping}}                                                                                                                                                        \\ \cline{2-5}   \hline
Attacker                         & \multicolumn{1}{p{3.5cm}|}{Source end adversary}                                                                                                 & \multicolumn{1}{p{3.5cm}|}{Channel end adversary}                                                                                                                & \multicolumn{1}{p{3.5cm}|}{Warden}                                                    & \multicolumn{1}{p{3.5cm}|}{Eavesdropper}                                                                                            \\ \hline
Goal                     & \multicolumn{2}{p{7cm}|}{\begin{tabular}[l]{@{}l@{}}1. Cause the receiver to get wrong information\\ 2. Tamper the results of downstream tasks\end{tabular}}                                                                       & {\begin{tabular}[l]{@{}p{3.5cm}@{}}Obtain the transmitter’s information\end{tabular}} & \begin{tabular}[l]{@{}p{3.5cm}@{}}Obtain the contents of source messages\end{tabular}                                                                \\ \hline
Knowledge                & \multicolumn{1}{p{3.5cm}|}{\begin{tabular}[l]{@{}l@{}}\emptycirc: Structure and weights\\ of semantic encoder \\ \fullcirc: Inputs and outputs of\\ semantic encoder\end{tabular}} & \multicolumn{1}{p{3.5cm}|}{\begin{tabular}[l]{@{}l@{}}\emptycirc: Structure and weights\\ of semantic codec\\ \fullcirc: Inputs and outputs of\\ semantic codec\end{tabular}} & {\begin{tabular}[c]{@{}p{3.5cm}@{}}~~~~~~~~~~~~~---\end{tabular}}  & \begin{tabular}[l]{@{}p{3.5cm}@{}}Structure and weights of semantic decoder\end{tabular} \\ \hline
Action                   & \multicolumn{1}{p{3.5cm}|}{Add noise into the source}                                                                 & Add noise into the channel                                                                                        & \multicolumn{1}{p{3.5cm}|}{Detect the occurrence of transmitting activities}                               & Decode the semantic information                                                                         \\ \hline
Defense                  & \multicolumn{1}{p{3.5cm}|}{Adversarial training~\cite{SemanticNoise}}                                                                                       & Adversarial training~\cite{Physical}                                                                                                       & \multicolumn{1}{p{3.5cm}|}{Covert communication~\cite{covert}}                                      & Encryption~\cite{evestrop}                                                                                 \\ \hline
\end{tabular}
\end{table*}
\section{Mixture-of-Experts (MoE)-based Semantic Communication for Trustworthy 6G Networks}

Although existing research works have each proposed a defense scheme to prevent a certain attack in SemCom systems, these approaches cannot simultaneously defend against multiple types of malicious attacks according to users' specific security requirements during communications, e.g., covertness and robustness. This limitation is particularly problematic in real-world scenarios. Therefore, to enable the practical application of SemCom, it is imperative to design an effective SemCom system to preserve telecommunication security under multiple heterogeneous threats for trustworthy 6G networks.

\subsection{Mixture-of-Experts (MoE) Network Model} 
The MoE network typically involves a gating network and several specialized trained models (i.e., experts), which can perform diverse machine learning tasks by selectively activating a subset of experts according to model inputs~\cite{MoE}. This simple network architecture caters to the demands for enhancing the trustworthiness of SemCom with a reasonable computational cost. In addition, the MoE can easily adapt to new datasets and AI tasks by adding or re-training experts without overhauling the entire network. With this scalable feature, an MoE-based SemCom system can retain normal operation by incorporating a newly trained expert whenever a novel security issue arises.

\subsection{An MoE-based Semantic Communication System for Trustworthy 6G Networks}

\begin{figure*}[t]
\centering{\includegraphics[width=0.95\textwidth]{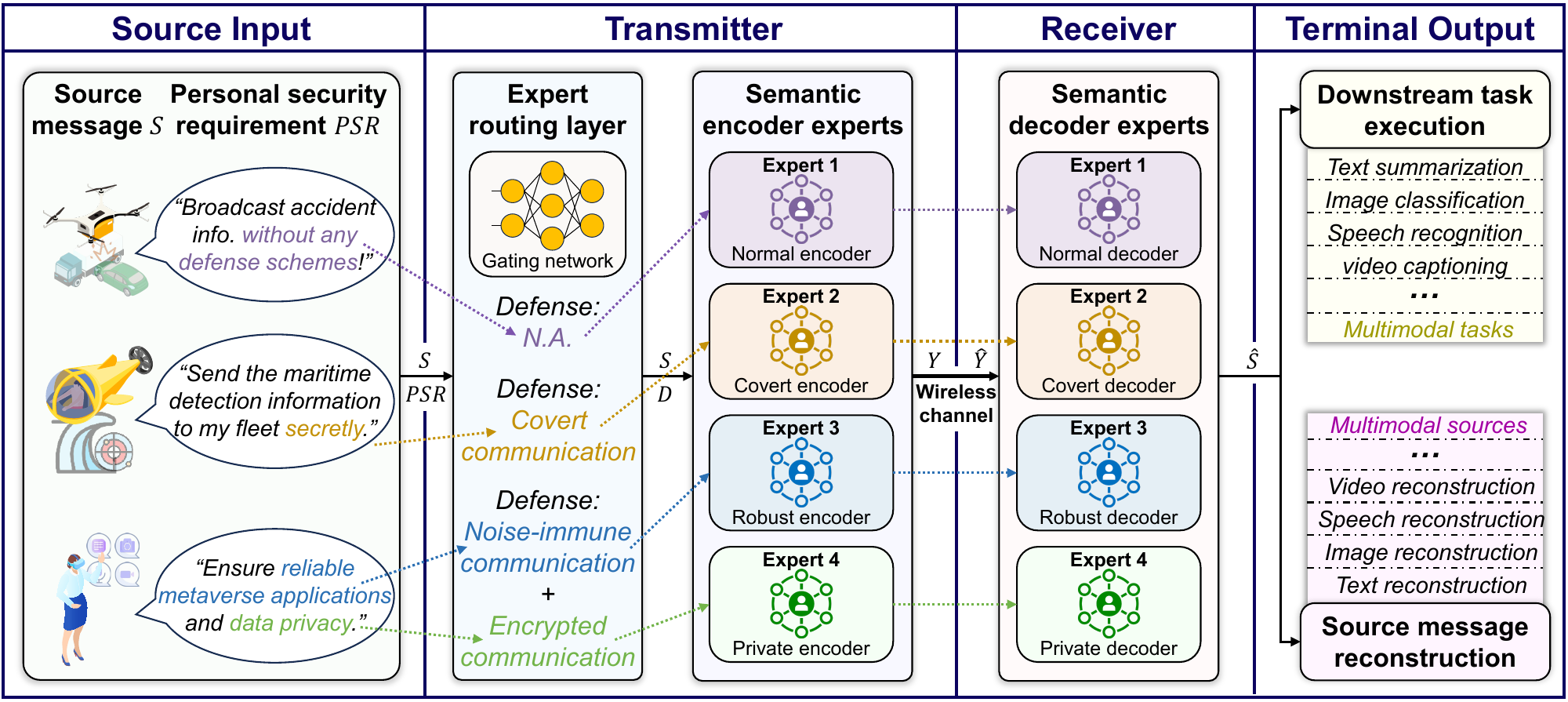}}
\caption{A general Mixture-of-Experts (MoE)-based Semantic Communication (SemCom) system building upon user-defined security requirements toward trustworthy 6G networks.}\label{Fig2}
\end{figure*}

As shown in Fig.~\ref{Fig2}, we propose an MoE-based SemCom system to handle users' diverse security requirements in 6G networks. The proposed SemCom system consists of a gating network, multiple semantic encoder experts for each modality at the transmitter end, and corresponding semantic decoder experts at the receiver end. The process of handling the source message and requirements in the MoE-based SemCom system is described below.

\begin{itemize}
    \item \textit{\textbf{Step1:}} The transmitter takes the source message $S$ and the user's personal security requirements $PSR$ as inputs to the gating network. 
    \item \textit{\textbf{Step2:}} The gating network outputs the decision $D$ about the selection and execution order of the experts based on source information and requirements.
    \item \textit{\textbf{Step3:}} According to decision  $D$, the transmitter calls the corresponding experts (semantic encoders and defense methods) to encode the source message $S$ to obtain semantic information $Y$.
    \item \textit{\textbf{Step4:}} After channel encoding, wireless channel, and channel decoding, the receiver obtains semantic information $\hat{Y}$.
    \item \textit{\textbf{Step5:}} By calling the corresponding decoder experts, the receiver recovers the source message $\hat{S}$.
\end{itemize}

To satisfy various user-defined security requirements, the data inputs from users of the proposed MoE-based SemCom system involve source messages and users' personal security requirements. By analyzing user objectives from the personal security requirement, the gating network can orchestrate the selection of experts to conduct security-oriented or privacy-preserving semantic encoding and decoding during SemCom. For example, users seek to immersively experience metaverse applications without privacy leakage~\cite{VTmigration_privacy}. This demands that during message transmission, the sensitive sensory data inputs (i.e., the source message $S$) should be anti-tampering and encrypted (i.e., the personal security requirement $PSR$). Based on $PSR$ and $S$, the gating network makes decisions $D$ to sequentially call the trained robust encoder and privacy encoder to extract the semantic information $Y$ credibly~\cite{LLM_MoE}. The extracted semantic information is then processed as $\hat{Y}$ by the channel encoder, wireless channel, and channel decoder. Then the privacy decoder and robust encoder at the receiver process the $\hat{Y}$ to recover $\hat{S}$ for the output. Since both robust codecs and privacy codecs are invoked, the communication process avoids message tampering and eavesdropping. The output $\hat{S}$ can be applied in various tasks, such as source message reconstruction and downstream task execution~\cite{MultiModal}.

\subsection{Gating Network and Experts Architectures and Training Method}
\textit{1) Gating network:} The key component of the MoE network is the gating network, which routes the source message to the proper experts. Here, we deploy a MultiLayer Perceptron (MLP) at the transmitter as the gating network to select one or more semantic codec experts based on users' security requirements~\cite{MoE}. The user's security and privacy requirements are embedded as a vector $PSR$ as part of the input to the gating network, with the source message $S$ to be transferred constituting the other part. The output of the gating network is a vector representing the decision $D$. During training, we randomly generate a dataset of uniformly distributed user requirements, and each requirement corresponds to a correct expert selection as the label. A user may have multiple requirements at the same time, corresponding to multiple labels. The loss function of the gating network comprises the selection loss and selection penalty of semantic codec experts, guiding the parameter updates of the MLP. The selection loss ensures that the gating network correctly routes the input to specialized experts to enhance task accuracy and meet user-defined security requirements. The selection penalty ensures that the gating network does not select unnecessary experts, reducing computing costs. Note that the gating network is trained after the expert's training, and the experts' model parameters are frozen during the gating network training. Therefore, when new experts are added, only the gating network needs to be fine-tuned, eliminating the need to retrain the entire system~\cite{MoE}. This architecture provides greater scalability for the SemCom system.

\textit{2) Normal codecs:} The architecture of a normal codec consists of a semantic encoder, a semantic decoder, a channel encoder, and a channel decoder~\cite{DeepSC}. The semantic and channel encoder is deployed at the transmitter end, while the semantic and channel decoder is deployed at the receiver end. The structure of normal codecs is different for different communication tasks. For example, for text transfer tasks, the Transformer model is used as a semantic codec~\cite{DeepSC}. For image transmission and classification tasks, Resnet is used as a semantic codec~\cite{SemanticNoise}. Regardless of the specific structure of the codec, the transmitter first feeds the $S$ into the semantic encoder to extract the semantic information $Y$. $Y$ is then sent to the receiver after passing through the channel encoder. Upon receiving the information, the receiver performs channel decoding to recover the semantic information $\hat{Y}$ and semantic decoding to recover the source message $\hat{S}$. By iteratively updating the model parameters of the semantic codec and channel codec based on the communication loss, we can train a normal codec expert.

\textit{3) Covert codec:} In traditional communications, a friendly jammer is usually set up to send a jamming signal to the sender to prevent the warden from detecting that communication is taking place. In SemCom, the semantic compression rate can be increased to improve communication concealment~\cite{covert}. The warden detects communication activities in the channel at a certain frequency. If a communication endures for an extended duration, each detection it encounters increases the likelihood of successful detection by the warden. Increasing the compression rate can reduce the number of detection times that the communication undergoes, thereby reducing the warden's detection success rate and achieving covert communication. Therefore, after obtaining the semantic information extracted by the semantic encoder, the covert codec further compresses it using a compressor. We utilize an autoencoder with an adjustable compression ratio as the compressor, where the encoder is deployed after the semantic encoder and the decoder is deployed before the semantic decoder. In the training phase, the autoencoder updates its parameters using the Mean Square Error (MSE) between the input and output as the loss.

\textit{4) Robust codec:} The Semantic Distance Minimization (SDM)-based training method is often used to combat source-end tampering attacks~\cite{SemanticNoise}. SDM uses adversarial training to reduce the distance between the semantic information of identical categories of images as much as possible. During training, SDM generates an online adversarial example for each benign example by adding semantic noise while updating the model parameters to make their semantic distance closer. Therefore, the semantic codec is largely immune to semantic noise. Although attacks from the source end and channel end are launched from different locations and in different ways, they both achieve the purpose of tampering with communication content by adding semantic noise. Therefore, the SDM is effective against tampering attacks from both the source end and the channel end.

\textit{5) Private codec:} The effective way to prevent information from being intercepted by eavesdroppers is to encrypt it~\cite{evestrop}. The transmitter and receiver each hold a secret key, with the sender concatenating the information to be transmitted with the key before semantic encoding. The legitimate receiver uses the key and semantic decoder to correctly recover the information, while malicious eavesdroppers, lacking the corresponding key, cannot recover the information even if they possess the same semantic decoder as the receiver. The privacy codec adopts adversarial training, simulating an eavesdropper holding the same semantic decoder, a pair of legitimate sender and receiver holding a key pair. The legitimate sender-receiver updates the model parameters to minimize the legitimate receiver's loss, while the simulated eavesdropper also attempts to minimize its loss. However, due to the advantage of the legitimate receiver holding the key, its decoder model gradually can converge, while the simulated eavesdropper does not.

In this section, we detail the architecture, training methods, and advantages of the proposed MoE-based SemCom system. However, calling multiple defense schemes simultaneously may result in a loss of communication accuracy. To explore its effects in real-world scenarios, we present a case study in the next section to evaluate its performance in a vehicular SemCom system.

 \section{Case Study}
\subsection{System Model}
As a case study, a simulation is conducted to assess the performance of the proposed MoE-based SemCom system. In this scenario, moving vehicles utilize semantic communication to exchange images captured by vehicular sensors, supporting applications such as autonomous driving and vehicular metaverses~\cite{SemanticNoise}. The environment includes different types of attackers attempting to tamper with information or eavesdrop on user privacy. The proposed MoE-based SemCom system, comprising a gating network, semantic codecs, and defense mechanisms, is deployed in the vehicles to counter these heterogeneous attacks. Based on this, vehicular users can customize their own security and privacy requirements based on the contents being transmitted.

We consider a scenario in which vehicles perform image task-oriented semantic communications with each other. The sender uses the MoE-based SemCom system to encode and send the acquired image, while the receiver decodes it and completes classification tasks~\cite{SemanticNoise}. During the communication process, we implement the four attacks mentioned above. The proposed SemCom system includes corresponding experts to resist the attacks and a gating network to precisely call these experts. Since the communication task is classification, a Resnet20 network is used as the normal codec~\cite{SemanticNoise}. The loss function for training normal codecs uses cross entropy as a classification loss.

\begin{figure*}[h] 
	\centering  
	\vspace{-0.35cm} 
	\subfigtopskip=2pt 
	\subfigbottomskip=2pt 
	\subfigcapskip=-5pt 
	\subfigure[Tampering attack from source end]{
		\label{fig3.a}
		\includegraphics[width=0.22\linewidth]{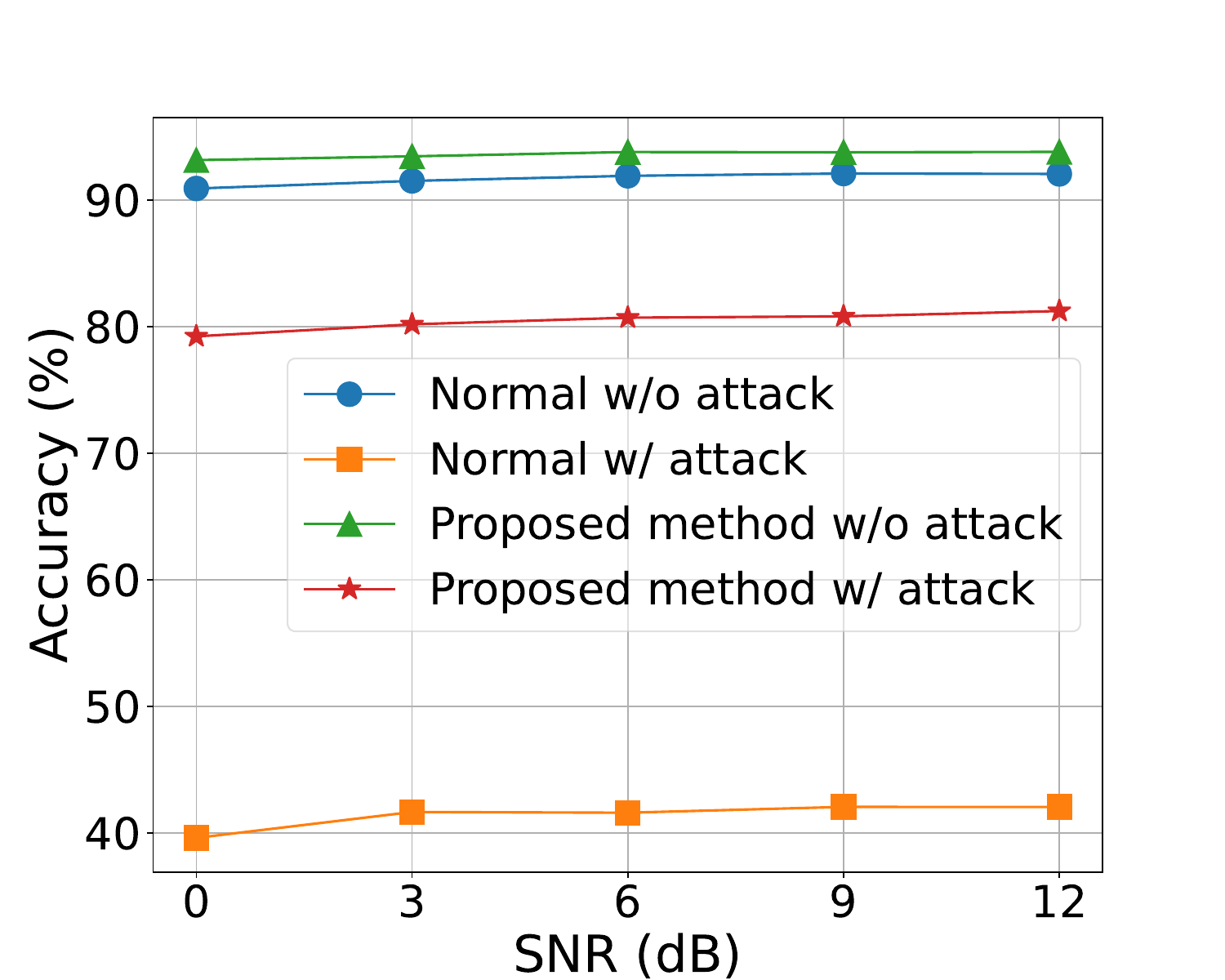}}
        \subfigure[Tampering attack from channel end]{
		\label{fig3.b}
		\includegraphics[width=0.22\linewidth]{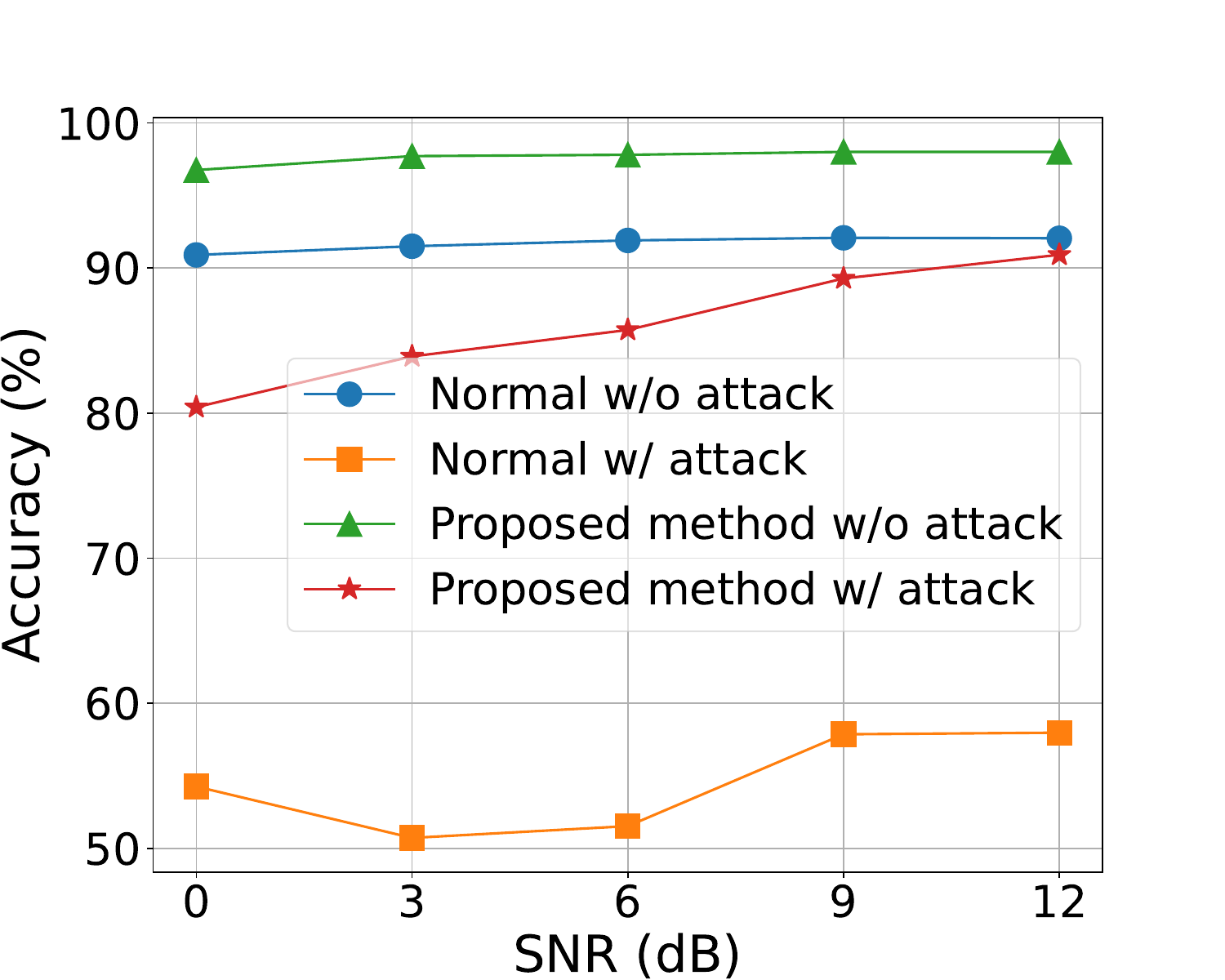}}
        \subfigure[Eavesdropping attack by a warden (SNR=12)]{
		\label{fig3.c}
		\includegraphics[width=0.22\linewidth]{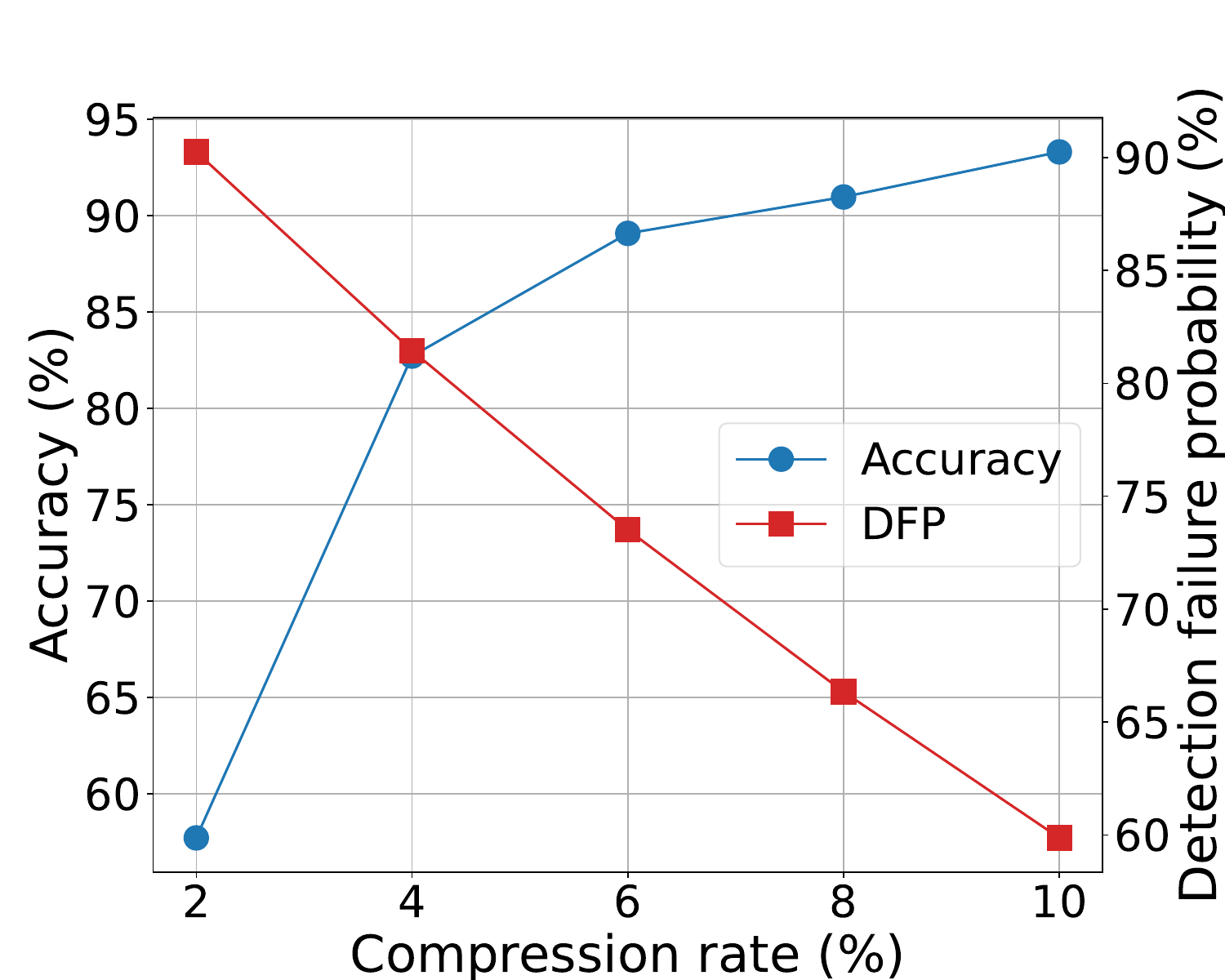}}
	\subfigure[Eavesdropping attack by an eavesdropper]{
		\label{fig3.d}
		\includegraphics[width=0.22\linewidth]{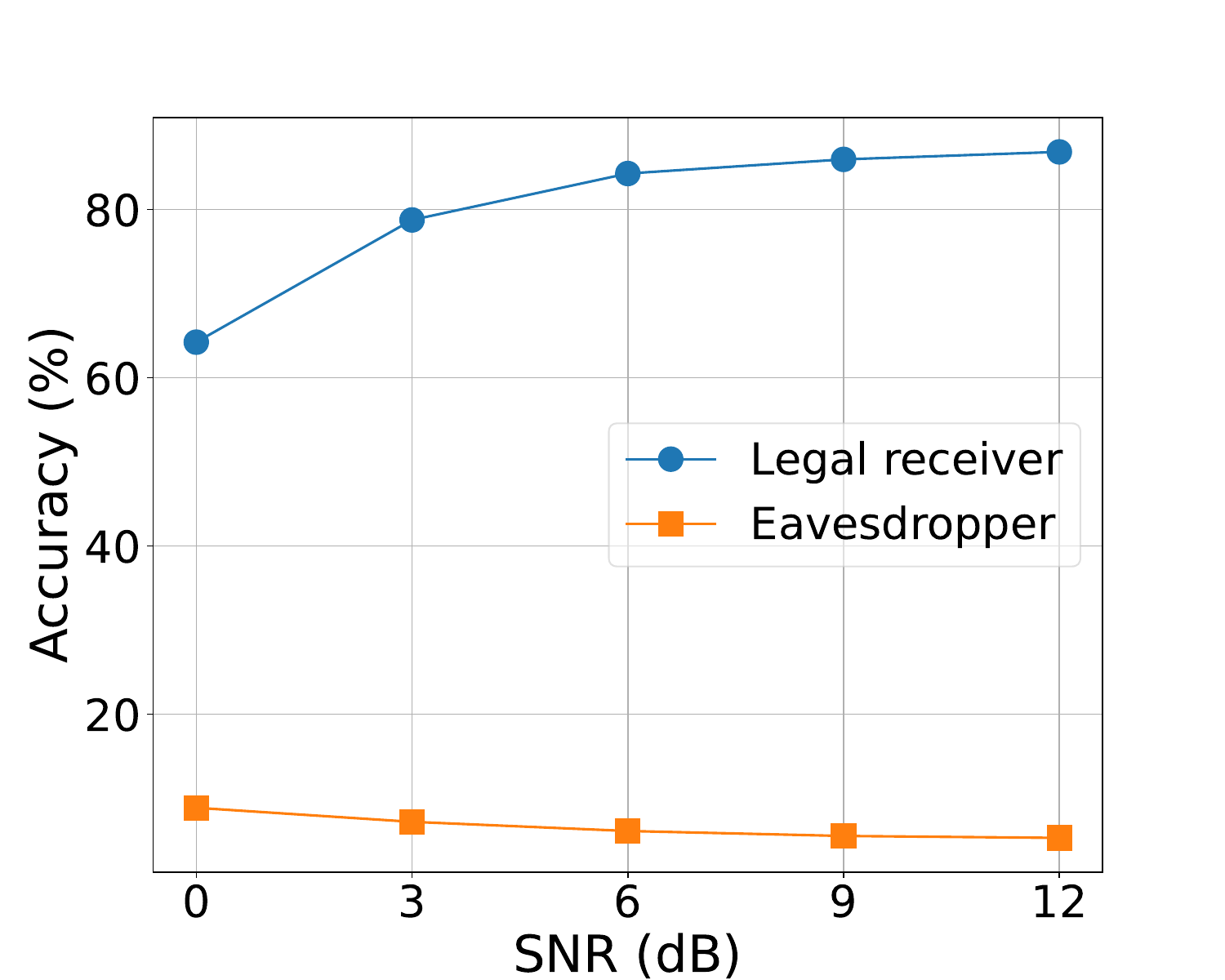}}
	\subfigure[Source end attack and warden]{
		\label{fig3.e}
		\includegraphics[width=0.22\linewidth]{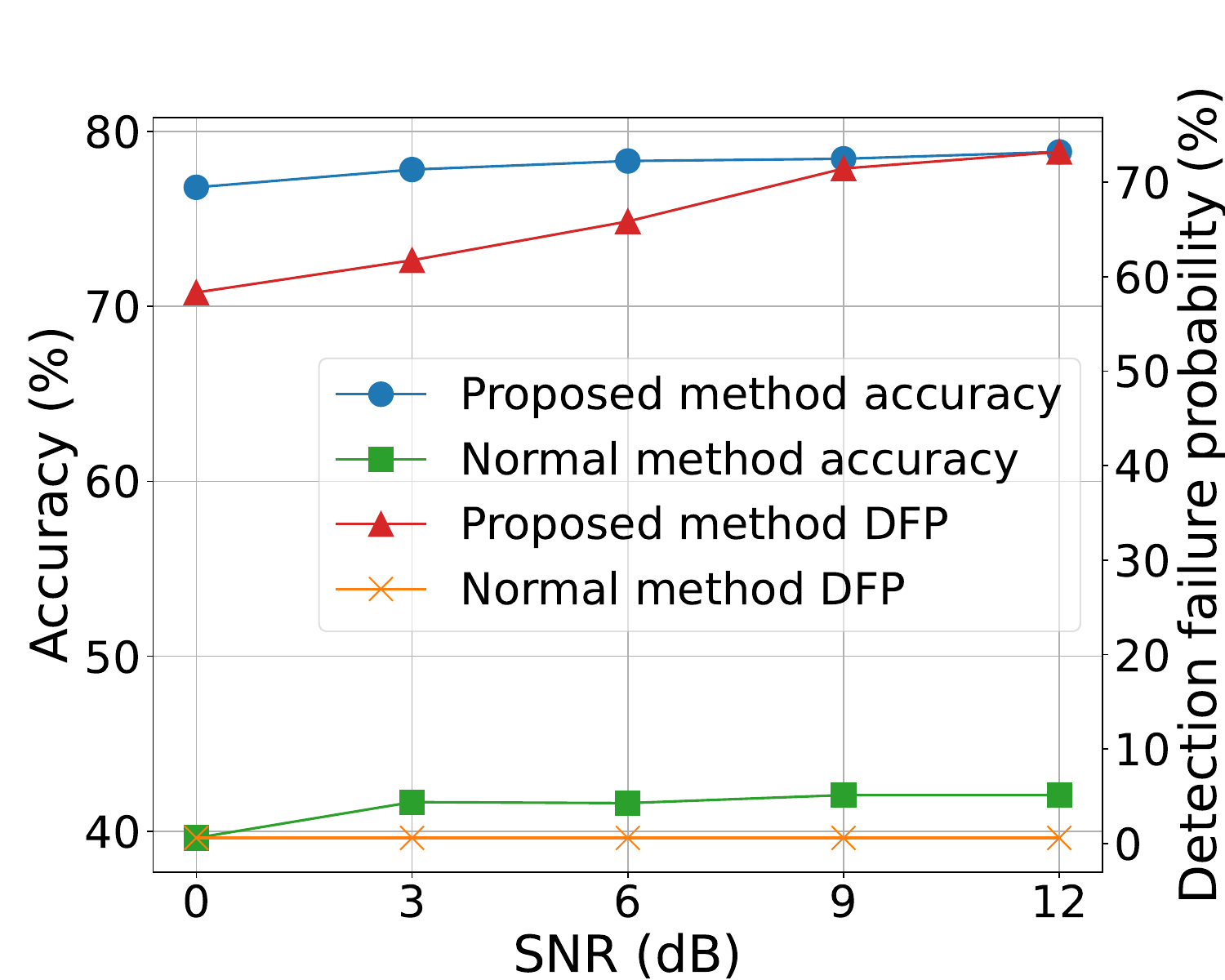}}
	\subfigure[Channel end attack and warden]{
		\label{fig3.f}
		\includegraphics[width=0.22\linewidth]{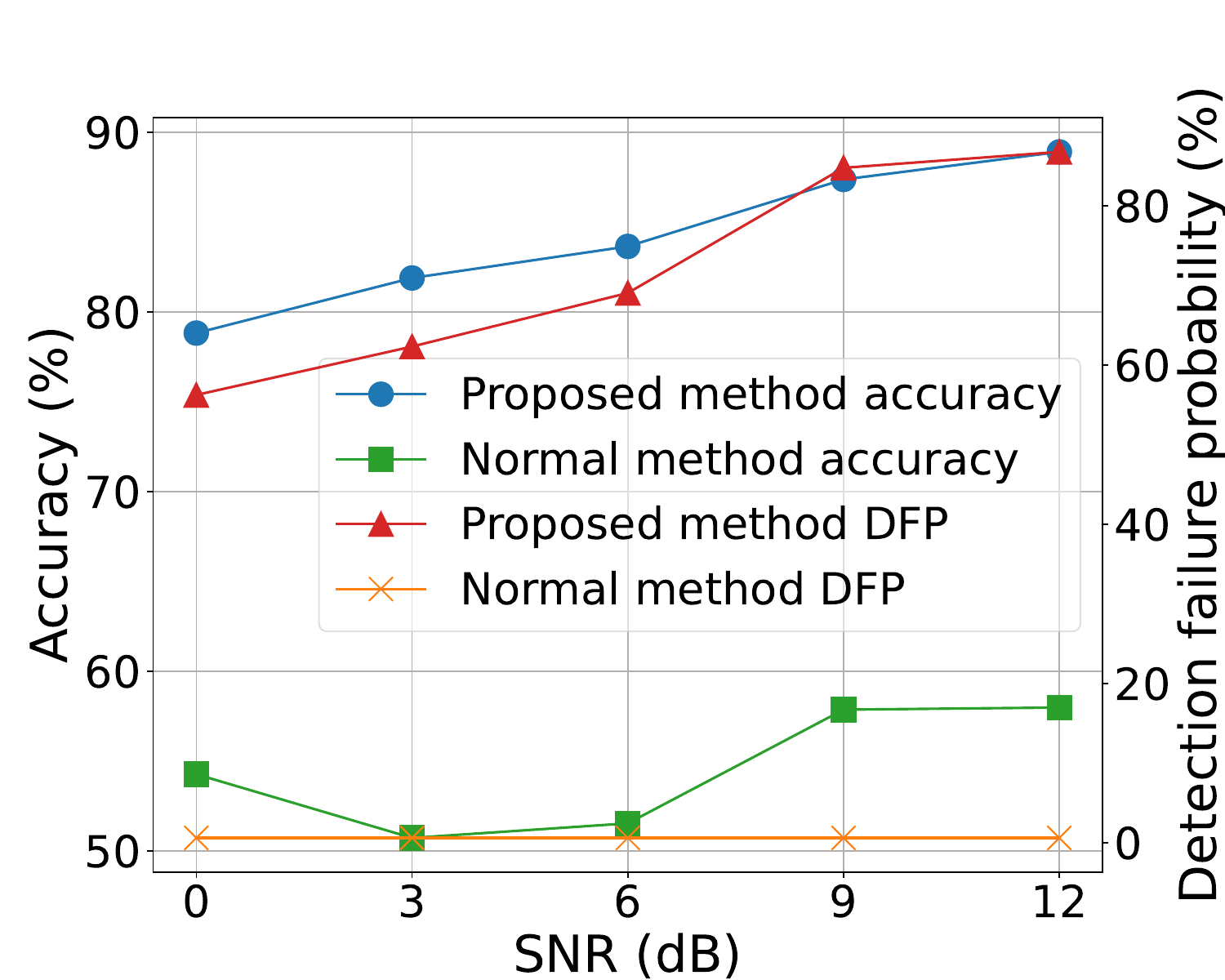}}
	\subfigure[warden and eavesdropper]{
		\label{fig3.g}
		\includegraphics[width=0.22\linewidth]{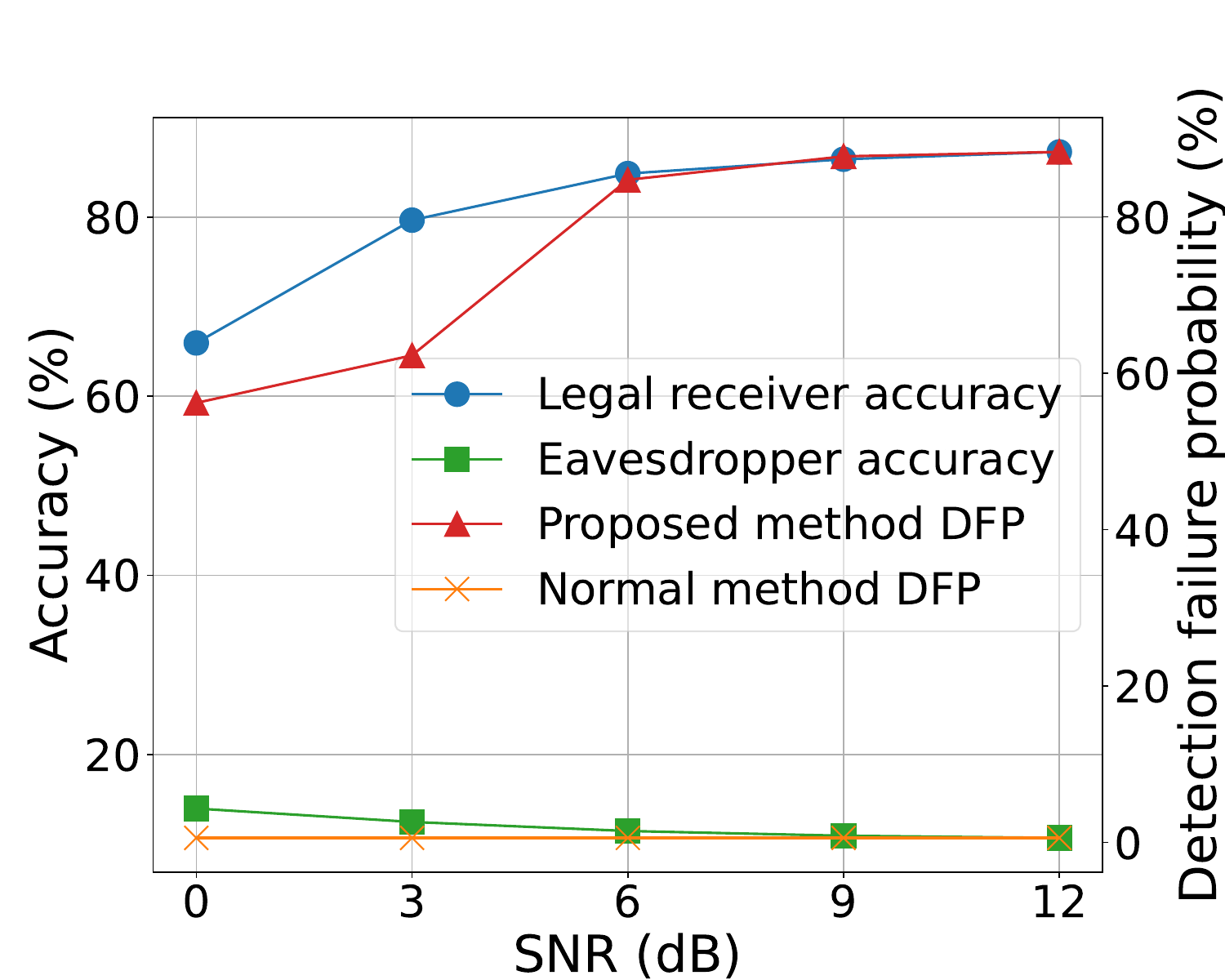}}
	\subfigure[Channel end attack and eavesdropper]{
		\label{fig3.h}
		\includegraphics[width=0.22\linewidth]{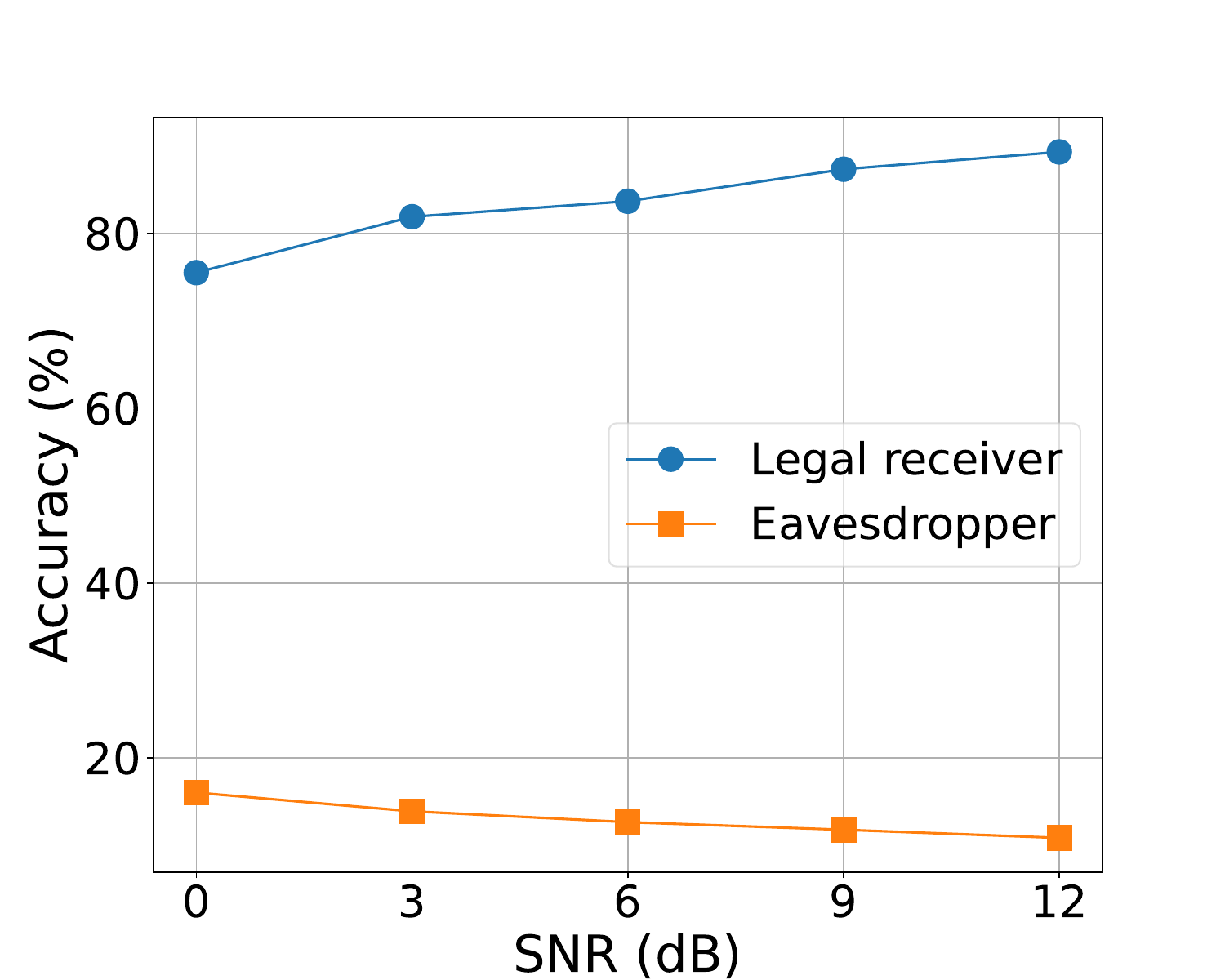}}
	\caption{The performance of proposed MoE-based SemCom system facing single and multiple heterogeneous attacks.}
	\label{fig3}
\end{figure*}
\subsection{Simulation Settings}
We consider a traffic sign recognition task to evaluate the performance of the proposed MoE-based SemCom system. The dataset used in the simulation experiments is the GTSRB ~$\footnote{https://benchmark.ini.rub.de/}$, which consists of over $50,000$ images with resolutions ranging from $15\times15$ to $250\times250$ pixels, divided into $43$ classes. 
During the performance evaluation of the MoE-based SemCom system, Additive White Gaussian Noise (AWGN) is added to the channel with a Signal Noise Ratio (SNR) between $0$ and $12$~\cite{DeepSC}. For the covert codec, the detection error probability, covert rate, and bandwidth are set to $95\%$, $100$ bit/s/Hz, and $5$ MHz, respectively. Besides, the warden monitoring the open wireless channel is assumed to detect communication activity at a frequency of twice per second~\cite{covert}.

The primary performance metric for a task-oriented SemCom system is the classification task accuracy. Since the goal of tampering attacks is to tamper with the results of classification tasks, classification accuracy is also used to measure the effect of tampering attacks. In particular, when the system suffers from an eavesdropping attack by a warden, the Detection Failure Probability (DFP) ~\cite{covert} metric is utilized to assess the effectiveness of covert communication in our proposed method. Meanwhile, the accuracy of an eavesdropper is also considered to evaluate the efficacy of semantic information encryption in the proposed system when facing an eavesdropping attack by an eavesdropper.


\subsection{Simulation Results}

Figure~\ref{fig3} shows the performance of the proposed MoE-based SemCom system when the system suffers attacks. The subfigures~\ref{fig3.a},~\ref{fig3.b},~\ref{fig3.c}, and~\ref{fig3.d} show the situation when it suffers a single attack, while the subfigures ~\ref{fig3.e},~\ref{fig3.f},~\ref{fig3.g}, and~\ref{fig3.h} are the case of multiple heterogeneous attacks. Various performance indicators prove that the proposed MoE-based SemCom system can adaptively select appropriate experts to complete communication tasks based on the user's security and privacy requirements. Numerical results show that the MoE-based SemCom system can effectively resist single attacks with minimal accuracy loss. 

Figure~\ref{fig3.a} shows the situation where SemCom is subject to tampering attacks from the source end. The MoE-based SemCom maintains a communication accuracy of about $80\%$ under various SNR ratio conditions, while the accuracy of the normal method drops to about $40\%$. In addition, even in the absence of attack interference, the MOE-based approach still has higher accuracy compared to the normal method. Similarly, as shown in Figure~\ref{fig3.b}, in the face of tampering attacks from the channel end, the MoE-based method can maintain an accuracy of more than $80\%$ and increase to $90\%$ with the improvement of channel quality. The accuracy of the general method falls below $60\%$ in the face of the attack from the channel end. It is worth noting that the accuracy of the proposed method in Figure~\ref{fig3.a} is not the same as that in Figure~\ref{fig3.b} without attacks. This is because the different requirements of the user cause the gating network to select different experts to handle the content to be transmitted. Since no previous paper has proposed a specific scheme to increase the DFP of an attacker by increasing the semantic compression rate in semantic communication, Figure~\ref{fig3.c} explores the accuracy performance and the DFP when adding covert codec experts with different compression rates to the MoE-based system. The numerical results show that the increase of compression ratio can significantly increase DFP, but the cost is that the communication accuracy will be weakened. Figure~\ref{fig3.d} shows the performance of the proposed method in the face of an eavesdropper, where the legal receiver can obtain high communication accuracy, especially at a high SNR, while the eavesdropper can hardly decode the message content. 

Figure~\ref{fig3.e} illustrates the performance of the proposed approach against both source-end attackers and a warden, showing that the MoE-based method achieves both high communication accuracy and DFP. As shown in Figure~\ref{fig3.f}, the proposed approach is equally advantageous in the face of channel-end attackers and wardens. Moreover, the MOE-based method can achieve similar results when facing a single type of attack with a high SNR. Figure~\ref{fig3.g} shows the simultaneous attack of the warden and the eavesdropper, where the MoE-based approach makes the eavesdropper achieve a very low precision parallel while maintaining the accuracy of the legal receiver and the DFP compared to the normal approach. As shown in Figure~\ref{fig3.h}, the proposed method can maintain stable communication accuracy and confidentiality in the face of channel-end attackers and eavesdroppers. In general, because the expert selection accuracy of the gating network is slightly decreased when facing the diversified security and privacy needs of users, the overall performance of the system is decreased compared with that of a single attack. However, it solves multiple heterogeneous attack scenarios that have never been considered by existing methods.

\section{Future Directions}

\subsection{Input Simplification for MoE-based SemCom}

The proposed MoE-based SemCom requires users to input their security and privacy requirements to tailor decision-making processes accordingly. Some users anticipate that these requirements will be addressed unconsciously, requiring minimal explicit guidance. For instance, if source messages include location details, the gating network autonomously selects the privacy codec without explicit user directives. By enabling automated decision-making based on user inputs and content characteristics, such systems can adeptly address security requirements and other requirements, such as multimodal SemCom~\cite{MultiModal}.

\subsection{Channel Estimation for MoE-based SemCom}

Accurate channel estimation is crucial for maintaining the integrity of a SemCom system, particularly as error rates of core semantic meanings have a greater impact. As SemCom evolves from traditional bit-based communication, a prolonged phase of hybrid bit-semantic communication is foreseen. The amalgamation of various semantic codec specialists and the hybrid nature of bit-semantic exchanges present hurdles in accurately estimating Channel State Information (CSI). Consequently, the pursuit of an effective and reliable channel estimation technique tailored for MoE-based SemCom emerges as a crucial avenue for future exploration in trustworthy 6G networks.

\subsection{Resource Allocation for MoE-based SemCom} 

The utilization of SemCom in resource-limited environments like the Internet of Things (IoT) is prevalent due to its capacity to notably diminish communication overhead. In MoE-based SemCom, each additional expert amplifies computational demands, augments storage requirements, and extends communication latency. In certain scenarios, prioritizing reduced energy consumption may warrant relinquishing unnecessary security measures. Moreover, optimizing storage resources by relocating seldom-used experts to the network edge rather than locally presents an avenue for efficiency. Consequently, the investigation into resource allocation assumes significant importance. Exploring the optimal approach for balancing constrained resources and security prerequisites for individual devices emerges as a pertinent future direction. 
		
\section{Conclusion}

Considering the multi-source heterogeneous attacks in the wireless communication environment, this article advocates for integrating Mixture-of-Experts (MoE) into Semantic Communication (SemCom) to enhance the trustworthiness of 6G networks from the aspects of security and privacy. By incorporating MoE into SemCom, the gating network can dynamically cater to users' varying security and privacy requirements of users by allocating suitable semantic codecs and security measures for each communication instance. To illustrate the efficacy of this integrated system, a case study focusing on vehicular semantic communications with four specific security threats is conducted. The results highlight the effectiveness of our approach in bolstering security and privacy in 6G networks, thereby laying a strong foundation for trustworthy network infrastructures. Finally, the potential future contributions of MoE for SemCom in advancing 6G networks are discussed, providing insights into the continued impact of MoE in promoting future communication paradigms.

\bibliographystyle{ieeetr}
\bibliography{ref}

\end{document}